\documentclass[MSNbibl,nameyear,dvips]{arxstspdf}
\usepackage{flushend}
\usepackage{stfloats}
\usepackage{graphicx}

\volume{27}
\issue{2}
\pubyear{2012}
\firstpage{189}
\lastpage{192}
\doi{10.1214/12-STS376B} 
\referstodoi{10.1214/11-STS376}



\begin{document}
\begin{frontmatter}

\vspace*{6pt}
\title{Discussion of ``Statistical Modeling of Spatial
Extremes'' by A. C. Davison, S.~A.~Padoan and M. Ribatet}
\runtitle{Discussion}

\begin{aug}
\author{\fnms{Darmesah} \snm{Gabda}},
\author{\fnms{Ross} \snm{Towe}},
\author{\fnms{Jennifer} \snm{Wadsworth}\corref{}\ead[label=e1]{j.wadsworth@lancs.ac.uk }}
\and
\author{\fnms{Jonathan} \snm{Tawn}\ead[label=e2]{j.tawn@lancaster.ac.uk}}
\runauthor{D. Gabda et al.}

\address{Darmesah Gabda and Ross Towe are PhD students,
Jennifer Wadsworth is Research Associate, Jonathan Tawn is Professor,
Department of Mathematics and Statistics, Fylde College, Lancaster University,
Lancaster, LA1 4YF, UK \printead{e2}.}

\end{aug}

\vspace*{-2pt}

\end{frontmatter}

We congratulate the authors for producing such a~helpful and
comprehensive overview paper of a~ra\-pidly developing and important area.
The starting point for inference in spatial extreme value problems is
to identify which features of the process require modeling. In certain
applications, for example, in the generation of return period maps,
only the marginal behavior is of concern. For such applications, the
covariate hierarchical/latent variable models reviewed in Section~4 are
ideal. However, if there is any dependence in the process between
sites, then care needs to be taken when assessing the uncertainty in
the estimated marginal distribution; the composite likelihood
procedures detailed in Section~6.2 can also be exploited in this
context when one wishes to avoid assumptions on the form of the spatial
dependence. As the authors point out, however, if interest lies in
modeling the joint occurrence of extremes over a region, then the
dependence structure of the spatial process needs to be explicitly
modeled. The most widely used approach in such cases is to model the
process as a~max-stable process. Here we will explore the suitability
of this framework for modeling spatial extremes, since these processes
are quite restrictive in their assumptions. Specifically, all finite
\mbox{dimensional} distributions of a max-stable process are multivariate
extreme distributions. Even simply in the bivariate case, this
corresponds to the variables being exactly independent\vadjust{\goodbreak} or
asymptotically dependent. Consequently, the broad class of
asymptotically independent variables is precluded under the modeling
assumptions of max-stable processes.

First consider diagnostic testing for the process being max-stable.
From our experience we feel that in many applications, max-stable
processes are assumed to be appropriate without testing their
suitability for the data. A partial justification for this is that
often it is pointwise maxima that are being modeled, and so just as one
appeals to the marginal limit theory to justify fitting the GEV
distribution site-wise, it seems natural to appeal to the limit theory
for the dependence structure also. However, we would typically not fit
the GEV to the margins blindly, but look to assess its suitability
through diagnostics such as Q-Q plots. To our knowledge, there
currently are no diagnostics for testing if the process is max-stable,
and so this discussion contribution aims to provide such a test.

Suppose that $\{Y(\mathbf{x})\dvtx \mathbf{x}\in A\}$ is a max-stable
process on the
region $A$ with standard Gumbel marginal distributions for all $\mathbf{x}$.
This is simply attained through a pointwise log transformation of the
more commonly-assumed standard Fr\'{e}chet margins. As the process will
typically only be observed at a finite set of sites $\mathbf{x}_1,
\ldots
,\mathbf{x}_m$, then all that can be tested for in practice is that the
joint distribution of $\{Y(\mathbf{x}_j)\dvtx\break j \in\Delta\}$, where
$\Delta=\{
1, \ldots,m\}$, follows a multivariate extreme value distribution.
Then for any $D\subseteq\Delta$ we have that
the joint distribution function for $\{Y(\mathbf{x}_j)\dvtx\break j \in D\}$ is
\begin{eqnarray*}
G_{D}(\mathbf{y})=\exp[-V_D\{\exp(\mathbf{y})\}],
\end{eqnarray*}
where $V_D$ is the associated exponent measure; see Section~2.3. A key
property of $G_D$, due to max-sta\-bility,
is that the distribution of $Y_D=\max_{j\in D}Y(\mathbf{x}_j)$ is
\begin{eqnarray*}
H_{D}(y) =G_{D}(y\mathbf{1})= \exp[-\exp\{-(y-\mu_D)\}].
\end{eqnarray*}
This is a Gumbel distribution with location parameter $\mu_D=\log
V_D(\mathbf{1})$ where
$0\leq\mu_D\leq\log(|D|)$, due to bounds on the exponent measure.
It follows that $Z_D=Y_D-\mu_D$\vadjust{\goodbreak} is standard Gumbel for all $D
\subseteq\Delta$. The idea of the diagnostic is to pool values of
$Z_D$ over replicates of the max-stable process and over
all $D\subseteq\Delta$, of a particular cardinality $k=|D|$, and test
using a P-P plot whether the variables $Z_D$ follow a standard Gumbel
distribution. This enables an assessment of the $k$th dimensional
properties of the process. Here we look at $k=2,3,4,m$.

There are a few practical issues to address. As the value of $\mu_D$
is unknown for all $D$ with $|D|\geq2$, these parameters require
estimation. We used maximum likelihood to estimate $\mu_D$, based upon
replicate data for $Y_D$, and subject to the parameter constraints
above; this was found to give better estimates than moment methods.
There are ${m \choose|D|}$\vspace*{1,5pt} sets with size $|D|$ in the power set of
$\Delta$, thus, for $m$ large, it is computationally intensive to
determine an estimate for all possible $\mu_D$. As a result, we limit
ourselves to
a randomly selected sample of 500 possibilities for~$D$ with $|D|=k$.
Although we did not explore this choice in more detail, it would seem
reasonable not to examine subsets $D$ in which the included sites are
likely to be independent, since the ability to detect a departure from
max-stability will be limited in these cases. In order to determine the
variability of our estimates, we use a nonparametric bootstrap to
produce 95$\%$ confidence intervals. Bootstrap methods are required as
there is clearly dependence in the pooled data over different $D$. By
treating each replicate of the process as a block, and constructing
$Z_D$ through estimation of $\mu_D$ for each bootstrap sample, this
complicated dependence is naturally incorporated into the confidence intervals.

Three different simulated data sets were used to illustrate the
methods, with 1000 replicates of the variables generated on a regular
grid of sites over a~$10\times10$ unit square. The different
dependence structures used were as follows: a Smith model max-stable
process (with identity covariance matrix $\Omega$); a~multivariate
extreme value distribution with logistic dependence structure
(dependence parameter $\alpha=0.7$); and a Gaussian process with
exponential correlation function ($\lambda=0.7^{-1}$), pointwise
trans-\break formed to have standard Gumbel marginal distributions. The first
two examples are max-stable in their dependence structure, that is,
follow multivariate extreme value distributions at the sites on the
grid. The third example, having a Gaussian copula, is not max-stable.

\begin{figure}

\includegraphics{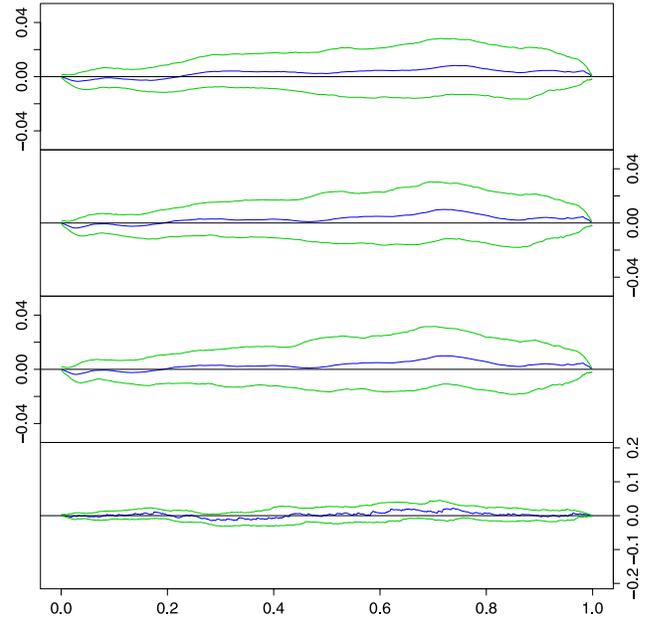}

\caption{Rescaled P-P plot for $Z_D$ derived from the Smith max-stable
process model. Top--bottom: $|D|=2,3,4,100$.}\label{Maxstable}
\end{figure}

\begin{figure}

\includegraphics{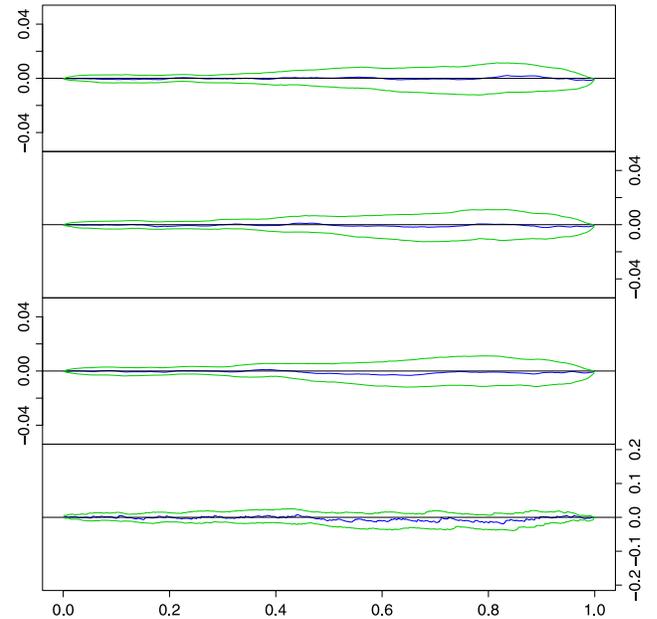}

\caption{As Figure~\protect\ref{Maxstable}, but for the multivariate logistic
distribution.}\label{mvevd}
\end{figure}

\begin{figure}

\includegraphics{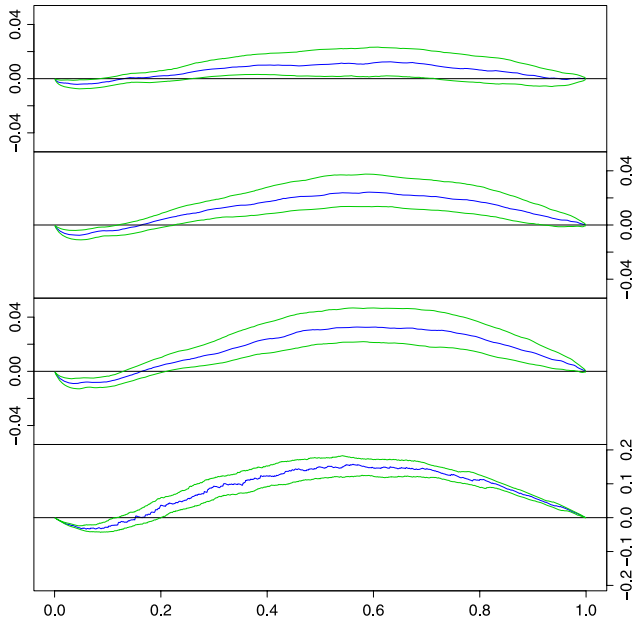}

\caption{As Figure~\protect\ref{Maxstable}, but for the Gaussian copula.}
\label{MVN}
\vspace*{-3pt}
\end{figure}

Figures \ref{Maxstable}--\ref{MVN} show the diagnostic P-P plots for the three dependence
structures respectively. Each figure illustrates the diagnostic for
$k=2,3,4,100$, with the P-P plot\vadjust{\goodbreak} presented on the $y$-axis as a
difference between the empirical and model probabilities. For the first
two processes the horizontal line, representing no departure from the
model probability, falls inside the pointwise 95\% confidence
intervals, thus indicating that the data provide no evidence against a
max-stable model. This is not the case for the Gaussian process copula,
which produces evidence to reject max-stability for each value of $k$.
Interestingly, the power of the test increases with $k$ (note the
change of scale on the bottom subplot). If only the $k=2$ case had been
used with a smaller sample size or weaker dependence we would wrongly
have failed to reject the null for the Gaussian copula. This indicates
that higher order dependence plays an important role in identifying the
nature of the extremal dependence. Currently, however, the standard
methods to fit max-stable processes use only the bivariate
distributions, through composite likelihood methods, as detailed in
Section~6.2.

More generally, this highlights that pairwise modeling techniques may
not always be as effective as one would hope. Even if the process has
bivariate extreme value pairwise distributions, this does not ensure
that higher order distributions are max-stable, or even asymptotically
dependent. The restriction to pairwise likelihood is to a large part
due to intractability of higher order distributions of max-stable
processes. As noted by the authors, higher order distributions for the
Smith model can be expressed (\cite{Genton11}), though the model's
utility is limited due to its unrealistic process realizations. In
terms of the copula models of Section~5, \citet{Nikoloulopoulos09}
provide the $d$-dimensional copulas for the extremal $t$ and, through
the appropriate limits, the H\"{u}sler--Reiss analogue. The authors
comment on the intriguing links between the copula and process models.
It would seem\vadjust{\goodbreak} to us that the extremal $t$ copula has the immediate
interpretation as the limiting finite-dimensional dependence structure
of normalized spatial $t$ processes. \citet{Kabluchko09} similarly
demonstrate that the Brown--Resnick, analogous to the H\"
{u}sler--Reiss, limit arises from normalized maxima of Gaussian
processes which become increasingly dependent through contraction of
the spatial domain at an appropriate rate. Consequently, these two
copula models could equally be viewed as max-stable process models. We
thus agree with the authors' suggestion that the connection is indeed a
matter of extending the copula to the full spatial domain. Evidently,
the utility of full spatial process models lies in being able to assess
features at unobserved sites; this is aided in the current context by
simulation of the full process. For the extremal $t$ model, the
definition-based method described in Section~7.1 and illustrated in
Figure~7 seems adequate for this. The alternative would be to derive
the appropriate spectral process $W(\mathbf{x})$ and exploit
representation~(20), though it is not clear that this would be simpler.
For the H\"{u}sler--Reiss extension, it appears that correspondence
with Brown--Resnick processes would permit simulation, through the
methods already employed in the paper.

At a more conceptual level, the major restriction of max-stable
processes is that the form of the extremal dependence at observed
levels must be assumed to hold for all more extreme events. However, it
may often seem more plausible that the dependence could weaken at
extreme levels, with the largest extremes becoming more isolated as
energy is concentrated into a more localized area. While max-stable
process models may well be sufficiently flexible to \emph{describe}
observed extremal dependence, for the purposes of \emph{extrapolation}
we would really like to be sure that the assumed stability holds. This
can also be examined through the empirical decay of tail probabilities.
Suppose $X(\mathbf{x})$ represents the original data process (before taking
pointwise maxima), but transformed to have common Gumbel margins.
Define $\bar{F}_D(x\mathbf{1})=\Pr\{X(\mathbf{x}_i)>x \dvtx i \in D\}$;
if the
dependence structure is asymptotically dependent, then $\bar
{F}_D(x\mathbf{1})= O\{\exp(-x)\}$ as $x\rightarrow\infty$. If this
feature is observed in the data, this suggests that the limiting
max-stable process is different from independence, and extrapolations
based upon fitting these processes may be reliable. If asymptotic
independence is present in the data, $\bar{F}_D(x\mathbf{1}) =O\{\exp
(-x/\break \eta_D)\}$, for some $\eta_D \in(0,1)$. If this feature is
observed in the data, then we would be rightfully skeptical~that
fitting a max-stable\vadjust{\goodbreak} process would provide reliable extrapolations.
Some recent work by \citet{Wadsworth12} explores the extremal
properties of a class of spatial processes which satisfy this
property.\vspace*{-3pt}

%

\end{document}